# THE GROUND STATE OF THE BOSE-HUBBARD MODEL IS A SUPERSOLID

ABSTRACT
The Bose-Hubbard model is well-defined description of a Bose solid which may be realistic for cold atoms in a periodic optical lattice. We show that contrary to accepted theories it can never have as a ground state a perfect "Mott insulator" solid and that it has a low-energy spectrum of vortex-like phase fluctuations. Whether the ground state is necessarily commensurate remains an open question.

Fisher et al[1] have introduced a model they describe as the "Bose-Hubbard" model for a quantum Bose solid, following my less explicit description of similar ideas[2]. This model has become of more interest because it can be made fairly realistic for bosonic atoms trapped in an optical lattice. The conclusion of reference 1 was that in a periodic structure with sufficiently strong repulsive interaction U the phase diagram as T→0 will consist, at almost all values of the chemical potential µ, of perfectly commensurate "insulating" solid, with only special, discrete values of µ allowing incommensurate, superfluid phases to exist. A fortiori, or so it seemed, there would also be no non-classical rotational inertia (NCRI).[3]

Recent experimental evidence[4] and theoretical arguments[5] have revived the old conjecture[6] that Bose solids at low T exhibit superflow. I shall show here that similar arguments apply in the Bose-Hubbard model and that a form of superfluidity, or at least of vorticity quantization, occurs for all µ. The Bose solid therefore stands in contrast to the Fermion case; but the question of whether the truly commensurate "Mott insulator" exists remains open.[7]

I will write the model Hamiltonian using the standard notation adapted from the Fermion case:

$$H = \sum_{i,j} t_{ij} b_i^* b_j + U \sum_i [n_i(n_i - 1) + (\varepsilon_0 - \mu) n_i] \quad [1]$$

Here i, j are sites on a lattice, $t_{ij}$ are hopping kinetic energy matrix elements, which are functions only of i-j and may be taken as non-zero only for near neighbors, and the $b_i^*$ and $b_i$ are respectively creation and destruction operators for bosons living on the sites of the lattice. One may think of the b's as bosons in site wave-functions $\varphi_i$:

$$b_i^* = \int d^D r \psi^*(r) \varphi_i(r) \quad [2]$$

where $\psi^*$ is the boson field. In the model the b's are taken to obey standard commutation relations

$$[b_i^*, b_j] = \delta_{ij} \quad [3]$$

which implies that the site functions $\varphi_i$ are orthonormal. One may think in terms of an energy band model whose spectrum is determined by the tight-binding coefficients $\varepsilon_0$ and $t_{ij}$; then the site functions $\varphi$ are the properly orthogonalized Wannier functions of this band. The ground state wave-function which is envisaged in reference 1 is the product function

$$\Psi = \prod_i b_i^* |vacuum\rangle \quad [4]$$

But [4] is manifestly not the true ground state because it has a finite matrix element $t_{ij}$ to any state in which site i is empty and a neighbor j is doubly occupied. This corresponds to the fact, emphasized in reference 4, that the true ground state of a many-boson system cannot change sign, and the orthogonalized Wannier functions of any simple band must do so. In order to arrive at a better approximation than [4] to the true ground state we must transform to non-orthogonal but more localized site functions $\varphi_i'$. That such a transformation is possible without going outside of the Hilbert space of a single band is known from the so-called "chemical pseudopotential" theory.[8]

I have relegated the simple but bulky calculations to an Appendix. Using the obvious fact that for small enough t/U the interactions between pairs of sites are simply additive, the calculations only involve two sites at a time and are quite easy. The interaction energy gained by mixing the site functions is, in this limit,

$\Delta E = - \sum_{pairs<ij>} t_{ij}^2 / U$ and the site bosons are

$$b_i' = N^{-1}[b_i + \sum_j t_{ij} b_j / U]; \quad N = \sqrt{1 + \sum \frac{t^2}{U^2}} \quad [5]$$

The reasoning of reference 6 is based on the theorem that a system is insulating for particles which are quanta of a conserved, complex field if and only if it has a local gauge symmetry in the phase of that complex field. The idea of this theorem is that if the energy is a function of the gradient of the local phase, $E = f(\nabla\varphi)$, $J = \partial E / \partial(\nabla\varphi)$, and a phase gradient is induced by a gradient of the chemical potential, a current will flow. If the ground state were correctly described by the naïve wave-function [4], the phases of the separate bosons would be meaningless, the number of particles on a site would not fluctuate, and the state would be a true "solid solid" as defined in ref 2. But if the bosons are not orthogonal, there are number fluctuations and it becomes a question whether E depends on $\nabla\varphi$. As we show in the Appendix, the mixing energy does depend on the relative phase and therefore the Bose-Hubbard solid is not an insulator at T=0.

As we see from the Appendix, the energy depends on the cosine of the relative phase, so it is reasonable to express the physics in terms of the supersolid model published first[9] in 2005. The effective hamiltonian for the phase degrees of freedom is the x-y model,

$$H_{eff} = -\sum_{ij} J_{ij} \cos(\varphi_i - \varphi_j) \quad \text{where} \quad J_{ij} \approx t^2/U \qquad [6]$$

There is every reason to believe that this result is general for Bose solids, although it is proved only for the simple case of the Hubbard model. For most such solids J will be so tiny as not to be observable, and phase fluctuations will restore rigidity. But it seems inescapable that there is a manifold of extra quantum phase degrees of freedom which is not included in the usual description of an elastic solid and which may manifest themselves at sufficiently low temperature. The dynamics of these degrees of freedom are best described in terms of the continuum approximation of ref [9]. A feature of the dynamics is that the phase fluid is modeled as

*incompressible*:
$$J = (\hbar/M)\nabla\varphi; \quad \nabla \cdot J \propto \nabla^2\varphi = 0$$
$$H = \hbar^2/M(\nabla\varphi)^2 \qquad [7]$$

so that the current does not affect the lattice of *sites*, necessarily. Here M is a parameter which can be quite large, having a ratio of order U/t to the physical helium mass. Of course, if [7] is valid everywhere, there is no real dynamic freedom; but under stress or at finite temperature point singularities (in 2D) or lines in 3D can exist; and their positions are the extra degrees of freedom with which the system responds. These singularities are surrounded by vortex flows which provide the greater part of their energy.[10]

I remark on the physical meaning of these extra degrees of dynamical freedom. The situation with respect to the Fermion and Boson solids has always seemed to me unsatisfactory. He3 has spin degrees of freedom and these undoubtedly order due to particle exchange at temperatures in the millidegree range for the bcc crystal. But until now there has been no corresponding exchange effect for the Bose solid. This seemed to me to be an anomalous asymmetry. The derivation of the coupling effect for the Bose-Hubbard has a close similarity to the derivation of "kinetic exchange" (superexchange)in the Mott insulator.

A second possible relationship is to the "TLS" degrees of freedom in glassy systems.[11] We might imagine that in the presence of disorder, the secular equation [A3]could have primarily localized solutions with a wide spectrum of energies extending through 0, which might serve as "two level centers". The observation of a distinct quantum system of degrees of freedom independent of the phonon spectrum is very parallel to our ideas.

The phase I am suggesting here may be distinct from a true superfluid. We can introduce the constraint $\nabla \cdot J = \hbar/M \nabla^2 \varphi = 0$ quite independently of the "supercurrent" definition $\delta E/\delta(\nabla \varphi) = \hbar J$. Only currents which flow *through* the sites would then exist: either uniform currents or vortex currents, and the only dynamics possible would be that of the vortex core singularities: a true "vortex fluid" such as Kubota[12] has advocated for He4. There are hints in the rather confusing data on solid He that this may be the nature of the NCRI in that case.

As far as cold atom experiments are concerned, the solidification transition into a nearly commensurate lattice of sites will still be observed. But if this lattice is rotated, it may show evidence of quantization of vorticity, especially for relatively large tunnelling. It will be a fascinating experiment to change the site-site interaction t gradually relative to U, preferably also studying the response to rotation.

Appendix

Here we present a calculation of the energy shift of two neighboring sites 1 and 2 in our supposed Bose-Hubbard model very similar to one carried out in ref [2], except that we focus on the dependence on the relative phase of the bosons at the two sites. For simplicity we confine ourselves to the case t/U<<1 where all site-site interaction energies are simply additive. In this limit it suffices to calculate that energy for a single pair. For a single pair the Hamiltonian is

$$H = t(b_1^* b_2 + b_2^* b_1) + U[n_1(n_1 - 1) + n_2(n_2 - 1)]$$
$$-\mu(n_1 + n_2) \quad [1]$$
$$\Psi_0 = b_1^* b_2^* |vac\rangle$$

What we want to do is to gradually apply a gauge field to the entire sample (or, equivalently, twist the boundary condition in such a way as to require a spiral phase rotation of the boson field). Thus each neighbor pair will pick up a relative phase difference $e^{i\eta}$ for the creation operator, $e^{-i\eta}$ for the destruction operator. The t term in [1] is manifestly not invariant to this change, but the U term is, and the naïve $\Psi_0$ presented in equation [1], which is an eigenfunction of the U term, is also. $\Psi_0$ is manifestly, however, not the ground state of the Hamiltonian [1]; our task is to see whether the true ground state shares this property.

It is easiest to calculate energies by the equation of motion method, avoiding the complications of normalization, which are considerable for bosons. If one creates a state by an operator O:

$$\Psi = O|vac\rangle,$$
$$[H, O]|vac\rangle = [E(\Psi) - E_0]\Psi \quad [2]$$

If [H,O] mixes different operators O', O'' etc, the resulting equations of motion must be diagonalized. First, calculating the EOM *without* the phase factor:

$$[t(b_1{}^*b_2 + b_2{}^*b_1), b_1{}^*b_2{}^*] = -t([b_1{}^*]^2 + [b_2{}^*]^2)$$
$$[U[n_1(n_1-1) + n_2(n_2-1)], b_1{}^*b_2{}^*\ ] = 0$$
$$[t(b_1{}^*b_2 + b_2{}^*b_1), ([b_1{}^*]^2 + [b_2{}^*]^2)] = -2tb_1{}^*b_2{}^*$$
$$[U[n_1(n_1-1) + n_2(n_2-1)], ([b_1{}^*]^2 + [b_2{}^*]^2) =$$
$$2U([b_1{}^*]^2 + [b_2{}^*]^2) \qquad [3]$$

The matrix to diagonalise is

$$\begin{pmatrix} 0 & -t \\ -2t & 2U \end{pmatrix} \text{ and } \lambda \approx -t^2/U$$

is the appropriate eigenvalue of the energy. The wave function is, to lowest order,

$$\Psi = [\ b_1{}^*b_2{}^* + t/U([b_1{}^*]^2 + [b_2{}^*]^2)$$
$$= b_1'{}^*b_2'{}^*, \text{ with } b_1' = ub_1 + vb_2 \text{ and } 1 \leftrightarrow 2,$$
$$u^2 + v^2 = 1, 2uv = t/U$$

[4]
As expected, the two b'*'s are not orthogonal.
Now we introduce the phase angle. It is essential that we realise that we are *not* solving the problem of an isolated pair of sites, but of a pair which we embed in an infinite or toroidally connected sample; we are giving the entire momentum spectrum a boost. Simplest is to consider the problem of a single site and its neighbors along the phase gradient A—which in a simple lattice will come in pairs on opposite sides of the 0 site. The boson belonging to such a site may be simplified to

$b_0' = ub_0 + v[b_1 + b_{-1}]$, and expanding in terms of
kinetic energy eigenbosons, [A5]
$$b_0' = N^{-1/2} \sum_k [u + 2v\cos k] b_k$$

The total kinetic energy is
$$E_K = -t/N \sum_k \cos k[u^2 + 4uv\cos k + v^2 \cos^2 k] \qquad [A6]$$
$$= -2tuv$$

Taking into account that $E_K$ is twice the energy gain, by the virial theorem, this agrees with [A3]. Now we ask how this energy depends on a boost of k by δk: cosk→cos(k+δk)=cosk cos δk −sink sin δk, and the second term averages out so the energy shift is just proportional to cos δk.